\documentclass[namedreferences]{SolarPhysics}
\usepackage[optionalrh,solaenum,natbib]{spr-sola-addons} 
\usepackage{graphicx}                    
\usepackage{color}                       
\usepackage{url}                         

\newcommand{\appr}{$\sim$}

\newcommand{\lapprox} {\, \lower3pt\hbox{\appr}\llap{\raise2pt\hbox{$<$}}\,}
\newcommand{\gapprox} {\, \lower3pt\hbox{\appr}\llap{\raise2pt\hbox{$>$}}\,}
\newcommand{\mycomment}[1]{}

\newcommand{\halpha}{H$\alpha$~}
\newcommand{\batse}{{\it CGRO/BATSE}}
\newcommand{\yohkohhxt}{{\it Yohkoh/HXT}}

\begin{document}

\begin{article}

\begin{opening}

\title{The RHESSI Microflare Height Distribution}

%
  \author{S.~\surname{Christe}$^{1}$\sep
  		S.~\surname{Krucker}$^{2,3}$\sep
		P.~\surname{Saint-Hilaire}$^{2}$
		}

%
\runningauthor{Christe et al.}
\runningtitle{The RHESSI Microflare Height Distribution}

%
\institute{$^{1}$ NASA Goddard Space Flight Center,Greenbelt, MD, 20771-0001, USA
                     email: \url{steven.d.christe@nasa.gov}\\ 
              $^{2}$ Space Sciences Laboratory, University of California at Berkeley, Berkeley, CA, 94720-7450, US
                     email: \url{skrucker@ssl.berkeley.edu} email: \url{shilaire@ssl.berkeley.edu}\\
              $^{3}$ University of Applied Sciences Northwestern Switzerland, Institute of 4D, Technologies, 5210 Windisch, Switzerland
             }

\begin{abstract}
We present the first in-depth statistical survey of flare source heights observed by RHESSI. 
Flares were found using a flare-finding algorithm designed to search the 6--10~keV count-rate when RHESSI's full sensitivity was available in order to find the smallest events \citep{christe2008micro}.  Between March 2002 and March 2007, a total of 25,006 events were found. Source locations were determined in the 4--10~keV, 10--15~keV, and 15--30~keV energy ranges for each event.  In order to extract the height distribution from the observed projected source positions, a forward-fit model was developed with an assumed source height distribution where height is 
measured from the photosphere.  We find that the best flare height distribution is given by $g(h) \propto
\mathrm{exp}(-h/\lambda)$ where $\lambda = 6.1\pm0.3$~Mm is the scale height.  A power-law height distribution with a negative power-law index, $\gamma = 3.1\pm0.1$ is also consistent with the data.  Interpreted as thermal loop top sources, these heights are compared to loops generated by a potential field model (\it{ PFSS}).  The measured flare heights distribution are found to be much steeper than the potential field loop height distribution which may be a signature of the flare energization process.
\end{abstract}

%

\end{opening}

%
\section{Introduction}
\label{sec:introduction}

Since the launch of {\it STEREO}, there has been renewed interest in understanding the three 
dimensional structure of solar flares.  Stereoscopic vision by the {\it STEREO}  
spacecraft, allow true source heights to be determined through
parallax.  Determining the height of solar flare sources is 
important in order to identify the location of the acceleration region and constrain flare 
models.  In the standard flare model, the electron acceleration region is situated in the corona above and produces closed magnetic loops which fill with hot thermal plasma observable in hard X-rays (HXR) up to 30~keV or more in large flares.  At the footpoints of these magnetic loops, accelerated particles are stopped by collisions and are observed down to 10 keV for the smallest flares.

Past studies have used a variety of different instruments and methods to determine the height of flare
sources on the Sun.  Using simultaneous observations of 607 flares by {\it Explorer 33, 35}, and {\it Mariner V}, \citet
{catalano1973} derived the height of soft X-ray sources (SXR, 1--6~keV) at the limb through the 
difference in occultation between observations with different view angles.  They found that 
their measurements of the dependence of emission on altitude above the photosphere 
was well fit by an exponential ($e^{-h/\lambda}$) with a scale height of $\lambda = 11\pm3$~Mm.  \citet
{kane1979} determined that a HXR source ($>$50~keV) was ``well 
below'' a height of \appr25~Mm using simultaneous observations by the {\it International Sun 
Earth Explorer 3} ({\it ISEE-3}) and the {\it Pioneer Venus Orbiter} ({\it PVO}).  Later, 
\citet{kane1983} showed that for impulsive HXR sources (150~keV)
95\% of emission originated at a height of $<$2.5~Mm above the 
photosphere.  Using {\it Hinotori}, \citet{ohki1983} studied two impulsive HXR (17--40~keV) 
bursts near the limb and found the height of the source centroid was $<$7~Mm.  
Later,  \citet{tsuneta1984} compared {\it Hinotori} and H$\alpha$ images and found 
the height of a HXR (16--38 keV) source to be \appr40~Mm.  \citet{takakura1986} compared HXR and \halpha 
positions and found a height of $7\pm3.5$~Mm for HXR (20--40~keV) sources.  
Using \yohkohhxt, \citet{matsushita1992} performed a statistical study comparing source locations in \halpha and HXRs 
(14--23~keV).  Interpreting the difference geometrically, she found an average 
source height of $9.7\pm2.0$~Mm and that the source height decreases as a function of energy.
The difference in heights were found to be $-1.0\pm0.3$~Mm, $-2.0\pm0.5$~Mm, $-3.2\pm0.7$~Mm in the four \yohkohhxt\
energy bands, 23--33~keV, 33--53~keV, 53--93~keV, respectively, relative to the lower energy channel.
\citet{asch1995} using \batse, investigated milliseconds time differences between peaks in emission in the 
25--50~keV and the 50--100~keV energy band.  Interpreted as a difference in the time of flight, \citet{asch1995} 
derived an altitude of $7.3\pm0.8$~Mm for the acceleration site.  Using RHESSI, \citet{asch2002} determined the absolute height of footpoints at different energies (from 15 to 50~keV) which occurred near the 
limb by assuming a thick-target model and a power-law density model.  Source heights 
between 4.5~Mm (at 15~keV) to 1~Mm (at 50~keV) were found.  More recently, \citet{sato2006} re-analyzed the {\it Yohkoh} flare
catalog using the method pioneered by \citet{matsushita1992}.  Comparing sources in the four \yohkohhxt\ 
energy bands, it was found that the difference in source heights between 
each energy band was $460\pm40$~km, $880\pm 90$~km, and $930\pm150$~km, respectively, smaller
than those found by \citet{matsushita1992}.  Recently, in two papers \citep{kontar2008, kontar2010}, Kontar et al. analyzed a GOES M6 class flare observed by RHESSI on the east solar limb.  They found observed source heights from 1 to 0.5~Mm for source from 20 to 200~keV consistent with thick-target emission.

Here, we consider the  RHESSI microflare height distribution.  
We use the RHESSI microflare list which is described and was previously analyzed in two 
papers by \citet{christe2008micro} and \citet{hannah2008micro}.   



\section{Observations}
\label{sec:observations}

RHESSI consists of 9 ultra-pure germanium detectors \citep{smith2002}, each behind a 
rotation modulation collimator (RMC) which enable imaging \citep{hurford2002}.  Each 
RMC is made up of a pair of identical grids of slits and slats with different pitches which 
modulate incoming X-rays as the spacecraft rotates.  Each RMC measures a different spatial 
frequency down to 2.3 arcseconds.  

RHESSI has been observing the Sun since its launch in February 2002.  A 
central product of the RHESSI collaboration is the so-called RHESSI flare 
list.  A separate RHESSI microflare list also exists
\citep{christe2008micro}.  This list contains the start, peak, end times, and 
background times along with positions of all microflares observed by RHESSI
from March 2002 to March 2007.  It contains 25,006 events of 
which 94.3\% have associated positions as found in the 6--12~keV energy range.  Small 
flare events with simple signatures of nonthermal electrons and thermal heating are well suited 
to study elementary flare processes compared to larger events with complex morphologies.  

\begin{figure}
\centerline{\includegraphics[width=0.9\textwidth,]{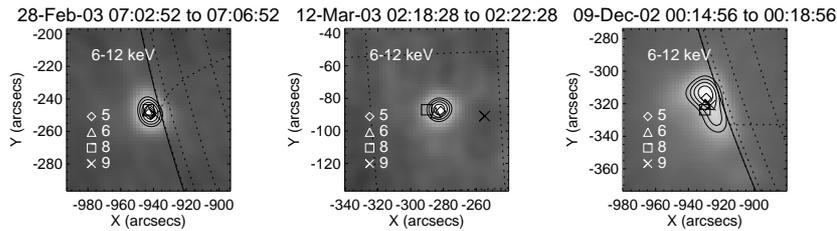}}
\caption{
Three examples of flare position finding.  Positions are found from back-projection images of each event integrated over 4 minutes around the peak of the flare (in the 6--12 keV band).  Images were created using detectors 5, 6, 8, and 9 individually (detector 7 has little sensitivity at low energies).  For each image, a paraboloid was fit to the maximum in the image.  Here, individual detector positions are shown overlaid on a representative back-projection image with 60\%, 70\%, 80\%, 90\% contours.  (\emph{Left}) An unresolved flare where positions from each detector agree.  (\emph{Center}) An unresolved flare where the flare position from detector 9 disagrees with the others due to the flare's proximity to the spin axis. (\emph{Right}) A resolved flare where positions from each detector disagree with each other since each detector resolves different sizes of the source.
}\label{fig:pos_finding} 
\end{figure}

Here, positions were redetermined for each microflare.  
Positions were found through back projection images of each event integrated over 4 minutes 
around the peak time in the energy ranges, 4--10, 10--15, and 15--30~keV.  The 
median nonthermal low energy cutoff for microflares was previously found to be 12~keV \citep
{hannah2008micro}, therefore the lowest energy band is dominated by thermal emission while the 
highest energy range is likely nonthermal.  Images were created using detectors 5, 6, 8, and 9 individually (detector 7 has little sensitivity at low energies and was excluded).  The angular resolutions are 20.4, 35.3, 105.8, 183.2 arcseconds, for detectors, 5, 6, 8, 9, respectively \citep{smith2002}.  For each image, a paraboloid was fit to the maximum in the image returning an interpolated flare position with errors (see Figure~\ref{fig:pos_finding} for examples).  

\begin{figure}
\centerline{\includegraphics[width=0.9\textwidth,]{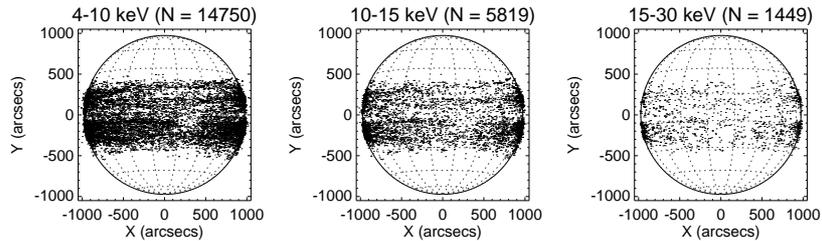}}
\caption{
The locations of RHESSI microflares from the RHESSI microflare list \citep{christe2008micro} re-evaluated 
using back projections images from detectors 5, 6, and 8 and averaging the positions from each.  Those events with a standard deviation in position greater than 5 arcseconds were excluded.  Flare positions are concentrated in the active region bands and  near the limb.  A lack of events can clearly be seen in the 4--10~keV positions around $[250, -100]$~arcsec, the average location of the spin axis, where imaging is difficult \citep{christe2008micro}. 
}\label{fig:positions} 
\end{figure}

The median flare loop length and width for microflares have been found to be 32 and 11 arcseconds, respectively \citep{hannah2008micro}.  This implies that detectors 6 and above may not resolve flare loops and should provide centroid positions that agree with each other.  {\it This fact provides a necessary cross check since it is not straightforward to algorithmically determine whether centroid locations which disagree between detectors are due to a resolved source or a noisy image}. In order to include only those flares with confirmed positions, results from detectors 5, 6, 8, and 9 were compared.  Those events with a standard deviation in position greater than 5 arcseconds (half of the median flare loop width) were excluded.  This reduced the number of events from 25,006 to 8,503 for the 4--10~keV energy range.  Detector 9 positions were found to differ significantly from other detectors.  This is likely due to the fact that detector 9 has the coarsest resolution and is therefore most often affected by the spin axis (where imaging is difficult).  The average location of the spin axis is around $[250, -100]$~arcsec \citep{christe2008micro}, therefore flares on the west limb are 4 times the resolution of detector 9 away from the spin axis and should be largely unaffected.  The final dataset now using detectors 5, 6, and 8 includes 14,750 events for 4--10~keV, 5,829 for 10--15~keV, and 1,452 for 15--30~keV.  The flare positions can be seen in Figure~\ref{fig:positions}.  Errors in the fitted centroids are small for all detectors.  The average error in flare positions for detector 5 is 0.68~arcsec.  These errors are smaller than the difference in positions between detectors, therefore the standard deviation of the set of positions is used as the error instead.

\section{Discussion}
\label{sec:discussion}

\begin{figure}
\centerline{\includegraphics[width=0.8\textwidth,]{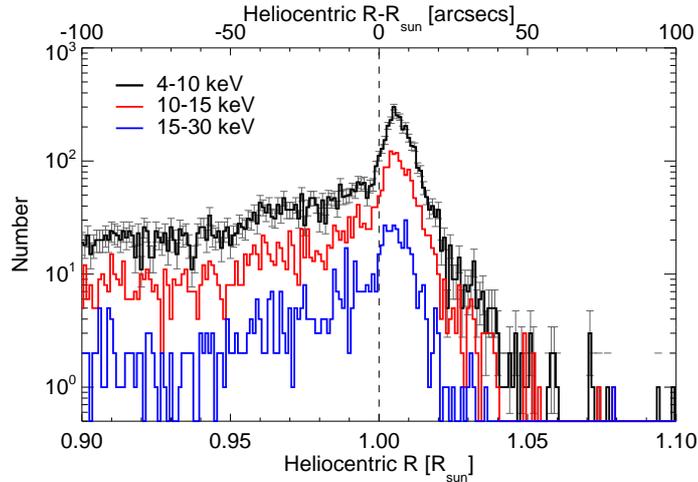}}
\caption{\label{fig:obs_distribution} 
The radial distribution of observed flare positions near the limb for each energy range (4--10~keV, 10--15~keV, 15--30~keV).  
The positions were corrected for the apparent size of the Sun and the solar B angle.  The dashed line represents the optical 
solar limb.  Positions are found to be concentrated at the limb with a peak at a heliocentric radius of approximately 1.005$R_{\odot}$ and a steeply falling distribution above the limb.  The peak in the distribution above the limb is an artifact due to partially-occulted sources.  
}
\end{figure}

Observations of solar source positions are a convolution of the radial source height and the spherical 
shape of the Sun.  It is therefore difficult to determine the absolute height above the photosphere for individual events.  In 
spherical coordinates, the relevant variables are the azimuthal angle, $\theta$ 
(or $\eta$, the longitude), the polar angle, $\phi$ ($\phi = \pi/2 - \delta$ where $
\delta$ is latitude), and the radius, $r$, which is related to the source height, $h$, by $r = 
R_{\odot} + h$ where $R_{\odot}$ is the solar radius.  The position of an event is fully 
described by $(r, \eta, \phi)$.  At the Earth, the only observable quantity is the two 
dimensional projection ($X = (R_\odot + h)\sin(\eta)\cos(\delta)$, $Y = (R_\odot + h)\sin(\delta)$) of this vector.  The height 
of a source is easiest to observe at the limb where projection effects are small.  In order to isolate the limb, we first define 
the radial distance of the source from the center of the solar disk as the projected source radius, $\rho = \sqrt{X^2 + Y^2}$, which is related to the coordinate variables through the following equation, 
\begin{equation}
	\rho = (R_\odot + h) \sqrt{\sin(\eta)^2\cos(\delta)^2 + \sin(\delta)^2}.
	\label{eq:rho}
\end{equation}
The measured projected source radius distribution for each energy band can be seen in Figure~
\ref{fig:obs_distribution}.  Most events are observed to occur above the limb.  

Generally none of the variables on the right-hand side of equation~\ref{eq:rho} are known.  
Yet if the distributions for each variable ($g(\eta), g
(\delta), g(h)$) is known then the distribution of the projected source radius, $g(\rho)$, 
can be determined.  Assuming all events are on the equator ($g(\delta) = 
0$), have identically zero height ($g(h) = 0$), and are uniformly distributed in longitude, the 
distribution function for $\rho$ is given by
\begin{equation}
	g(\rho) = \frac{A}{\sqrt{1-(\rho/R_{\odot})^2}}
	\label{eq:simplemodel}
\end{equation}
where A is a normalization constant.  Generally, deriving an analytical solution for $g
(\rho)$ is impossible, therefore a forward fit model was developed.  The model takes as input a source height distribution, $g(h)$, a longitude and latitude distribution and generates random vectors ($h$, $\eta$, $\delta$).  The 
longitude distribution is assumed to be uniform\footnote{It is incorrect to select 
longitude from a uniform distribution in order to have points distributed uniformly over a sphere 
since the spherical area element depends on latitude.  This effect is small (up to a few percent) since most 
flares occur near the solar equator and simulations show that it does not affect the results presented here.} while random latitudes are generated based on the observed microflare latitude distribution \citep[see][]{christe2008micro}.  These random 
vectors are then projected onto the two dimensional visible solar disk as ($X$, $Y$) excluding those that are occulted.  The source projected radius, $\rho$, is then calculated for each event and the distribution, $g(\rho)$, can be compared with the measured distribution to determine the height distribution, $g(h)$.  

In this study, two height model distributions are considered; an exponential distribution defined by $g(h) = A\ \mathrm
{exp}(\frac{-h}{\lambda})$ and a power-law distribution, $g(h) = Ah^{-\alpha}$.  
These two models were chosen for their mathematical simplicity following the example of 
\citet{catalano1973} and are not necessarily physically motivated.  

The discussion so far has assumed point sources.  Finite source size should have no effect on the 
projected radial distribution for sources well above the photosphere ($\rho>\sigma_{size}$).  
Those sources whose maxima are behind the limb yet are within \appr$\sigma_{size}$ will show 
emission truncated by the solar limb.  RHESSI imaging of a partially occulted source will be ``blurred'' and
show a source maximum above the limb whose height is dependent on the detector resolution.  This effect places a minimum
threshold below which the height of occulted sources cannot be accurately determined by RHESSI and is the reason for the maximum in the observed radial distribution.  For detector 5, 6, and, 8 the height of the peak in the radial distribution was found to be 5.0, 5.3, and 6.7 arcsec above the limb, respectively.  Height determination for sources as close to the limb as these values should be investigated carefully in order to determine whether they are occulted.  In the following analysis, both height models include an unphysical minimum cutoff height, $h_{0}$, to model this effect. Using only unresolved sources, as is done here, is unlikely to affect the shape of the height distribution though there may be an affect on the minimum source height value if the source height is related to the source size.

\section{Analysis}

\begin{figure}
\centerline{\hspace*{0.015\textwidth}
               \includegraphics[width=0.515\textwidth,]{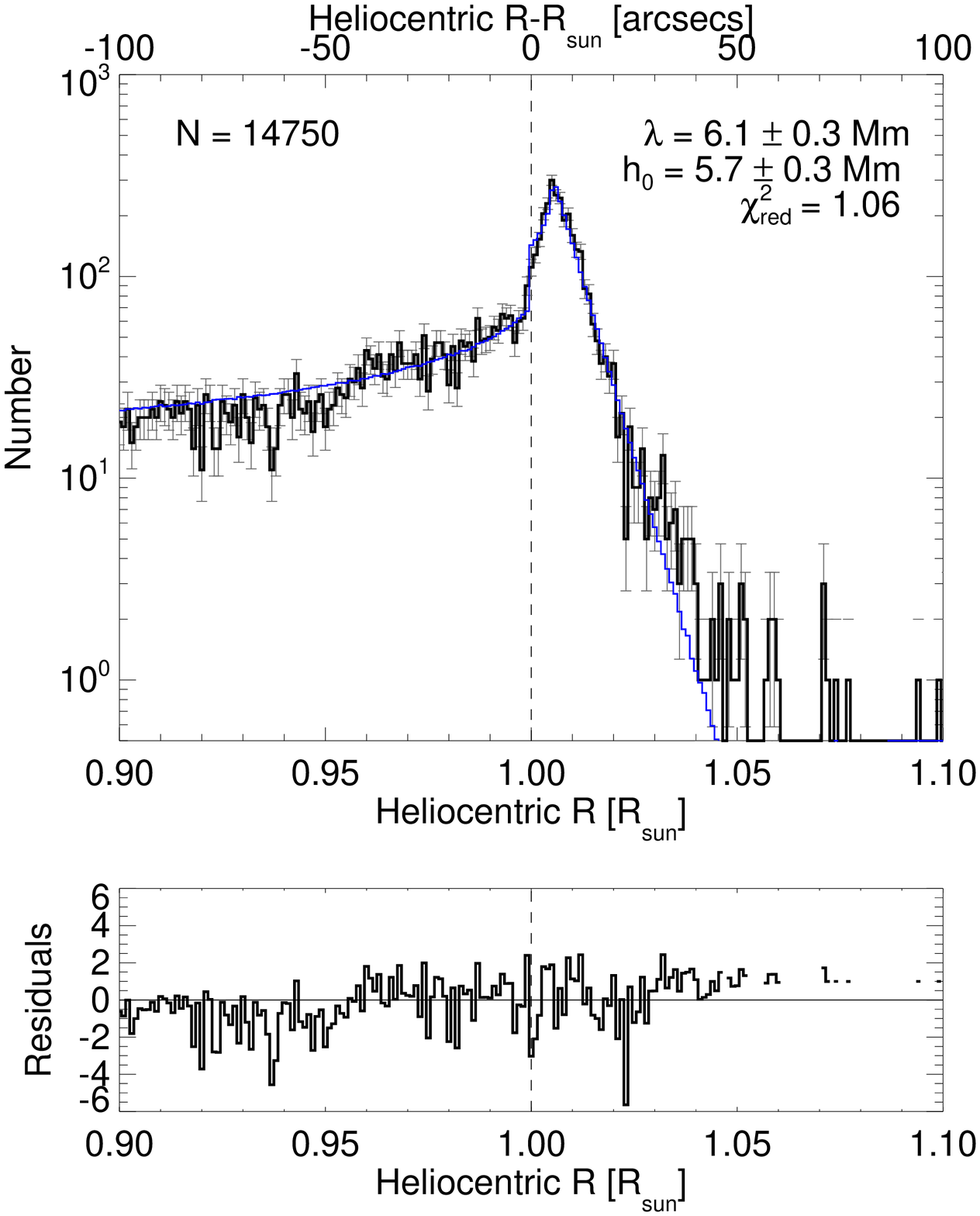}
               \hspace*{-0.03\textwidth}
               \includegraphics[width=0.515\textwidth,]{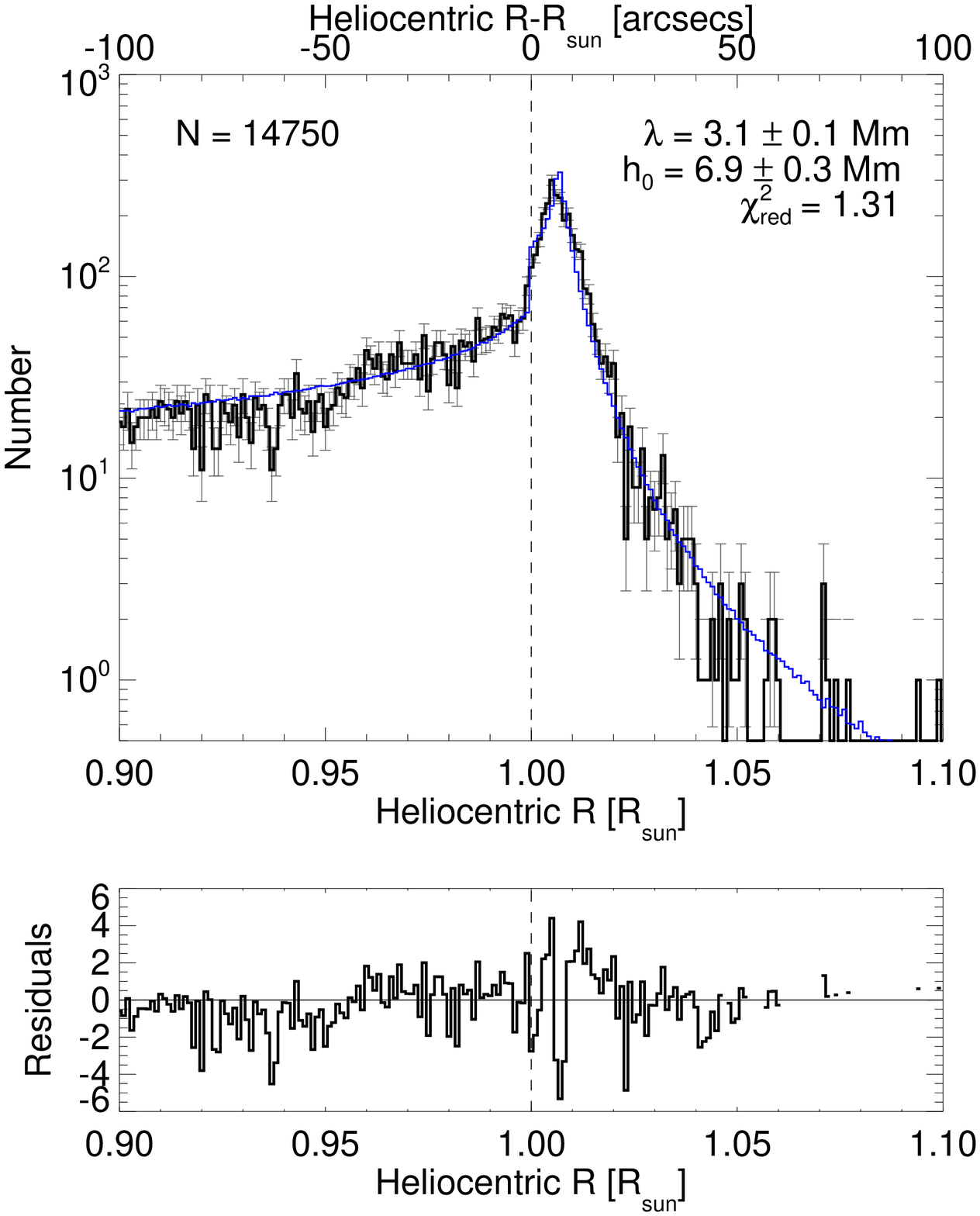}
}              
\caption{
The radial distribution of observed flare positions near the limb in the 4--10~keV (black) energy 
bands.  The dashed line represents the optical solar limb.  (\emph{Left}) The best fit assuming an exponential 
height distribution is overlaid in blue.  The scale and minimum height were found to be
$6.1\pm0.3$~Mm and $5.7\pm0.3$~Mm, respectively.  The exponential fit captures the overall behavior of the 
data though underestimates the number of events at large heliocentric radii.  (\emph{Right}) The best fit assuming a 
power-law height distribution.  The power-law index and minimum height were found to be $3.1\pm0.1$~Mm and 
$6.9\pm0.3$~Mm, respectively.  The power-law height distribution fit is worse than the exponential height 
distribution as measured by the reduced $\chi^{2}$ but better captures the events at large heliocentric radii.
}\label{fig:obs_fit} 
\end{figure}

Using the method described above, the height model distributions were fit to the observed distribution through multiple Monte 
Carlo runs.  Best fit parameters were determined through a grid search for the minimum in the $\chi^2$ space.  
The number of simulated events required for the Monte Carlo simulation were chosen in order to 
make the variance in the minimum smaller than the errors as determined by the contours 
of the $\chi^2$-space while maintaining an acceptable computation time.  For the exponential height model, in the 4--10 keV 
energy range, the scale height, $\lambda$, was found to be $6.1\pm0.3$~Mm and the minimum 
height, $h_{0}$, is $5.7\pm0.3$ with a reduced $\chi^{2}$ of $1.06$.  The fit results can be seen in Figure~\ref{fig:obs_fit}.  The authors would like to point out the presence of a step-like structure at $\approx0.96 R_{\odot}$.  This feature is statistically significant and is most prominent in the eastern hemisphere.  Its cause is unknown to the authors but it does not affect the results presented here.  The fit results disagree with those found by \citet{catalano1973} who found a scale height of $11\pm3$~Mm in the 1--6~keV range though \citet{asch2002} concluded that most past height measurements have large systematic systematic errors.  \halpha flare positions, which \citet{catalano1973} depends on, have an associated accuracy of 1 heliographic degree (or 12~Mm).  Analyzing the study by \citet{matsushita1992}, which similarly depend on $H\alpha$ positions, \citet{asch2002}  required a correction factor of $-6.8$~Mm which, if applied to \citet{catalano1973}, would lead to agreement with the results found here.  

Fits were also performed for the 10--15 keV and 15--30 keV energy range.  Though those results have 
poorer statistics, they agree with the 4--10~keV fit to within $1\sigma$.  These results are consistent with past observations of thermal loops such as those of \citet{ohki1983} and \citet{takakura1986}.  

The power-law height distribution was also fit to the data for each energy channel.  The power-law index, $\alpha$, was found 
to be $-3.1\pm0.1$ with a minimum height, $h_{0}$, of $6.9\pm0.3$~Mm, similar to the minimum height found in the
exponential height model, and a reduced $\chi^{2}$ of $1.31$.  Again, the results at the higher energies were found to agree.  Both the exponential and power-law height models fit the data well with a small reduced $\chi^{2}$.  The exponential fit was found to fit the data close to the limb better than the power-law model.  This fact explains the lower value of $\chi^{2}$ for the exponential fit.  On the other hand, the power-law fit better captures the tail of events at high heliocentric radius. 

\begin{figure}
  \centerline{\includegraphics[width=0.9\textwidth]{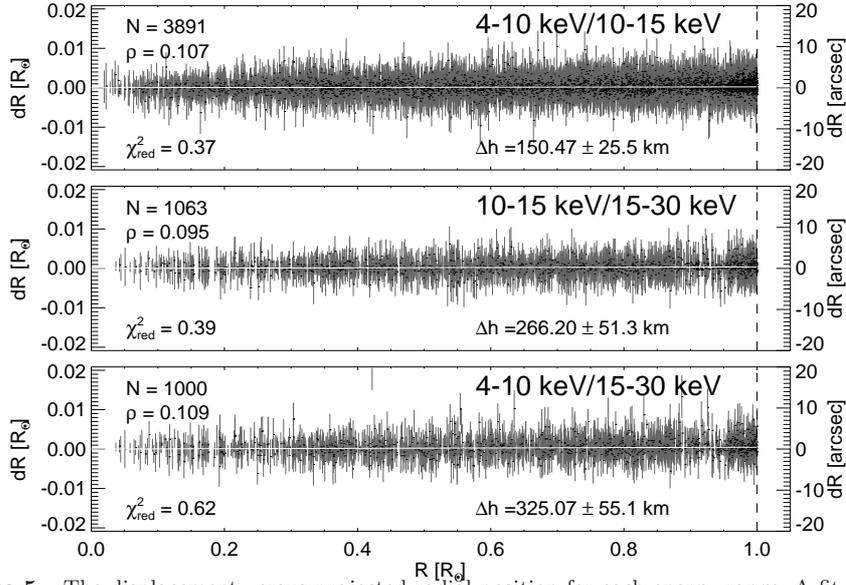}}
\caption{
The displacement versus projected radial position for each energy range.  A fit to the data yields the average height difference 
between energy ranges (white line).  The errors bars represent the standard deviation of the positions from different detectors.  The average height differences are small, $\Delta h_{4-10,10-15\ \mathrm{keV}} = 150\pm25$~km, $\Delta h_
{10-15,15-30\ \mathrm{keV}} = 270\pm51$~km, and $\Delta h_{4-10,15-30\ \mathrm{keV}} = 320\pm55$~km where lower 
energies are at larger heights.  The results are found to be self-consistent since $\Delta h_{4-10,15-30\ 
\mathrm{keV}} = \Delta h_{4-10,10-15\ \mathrm{keV}} + \Delta h_{10-15,15-30\ \mathrm{keV}}$.  
}\label{fig:matsushita} 
\end{figure}

To better compare the source heights between different energies, a method developed 
by \citet{matsushita1992} and more recently applied by \citet{sato2006} is appropriate here.  Assuming that 
multiple pairs of points lie on either one of two concentric spheres with different radii, then the relationship 
between the projected radial positions and the length of the displacement vectors between each pair of points is 
related to the difference in radius (or height, $h$) between the two spheres.  This method, by design, ignores those events observed above the limb and is therefore not sensitive to the problem of partially occulted sources.  Figure~\ref{fig:matsushita} displays this relationship for each energy range.  The errors bars represent the standard deviation of the positions from 
different detectors.  The fitted height differences were found to be small; $\Delta h_{4-10,10-15\ \mathrm{keV}} = 
150\pm25$~km, $\Delta h_{10-15,15-30\ \mathrm{keV}} = 270\pm51$~km, and $\Delta h_{4-10,15-30\ \mathrm{keV}} = 
320\pm55$~km where lower energies are at larger heights.  These results  
are found to be self-consistent since it must be that $\Delta h_{4-10,15-30\ \mathrm{keV}} = \Delta h_{4-10,10-15\ \mathrm{keV}} + \Delta h_{10-15,15-30\ \mathrm{keV}}$.  The reduced $\chi^{2}$ for all of the fits are less than one which suggests that the errors, which here represent the standard deviation in the set of detector positions, are too large.  This may indicate that source structure is being resolved.  Comparing to previous uses of this method, it is found that the 
height differences are small compared to results by both \citet{matsushita1992} and \citet{sato2006}.  The absolute heights in each energy band suggests that images in all energy bands are dominated by thermal sources.  

In order to better interpret the fitted height distribution, we used the potential-field source-surface ({\it PFSS}) to model
(non-flaring) magnetic loops.  This model calculates the magnetic field in the corona using photospheric magnetic field observations ({\it SOHO/MDI}) assuming that there are no currents ($\nabla \times B = 0$).  This model was initially developed by \citet{altschuler1969} and \citet{schatten1969}, and later refined by \citet{hoeksema1984} and \citet{wang1992}.  The {\it PFSS} model, through its assumption of zero current, models the lowest energy state of the (non-flaring) corona.  It is, therefore, not an accurate model of the actual corona below $\sim$1.6 solar radii yet it provides a useful theoretical ``base'' corona to compare with.

\begin{figure}
  \centerline{\includegraphics[width=0.8\textwidth,]{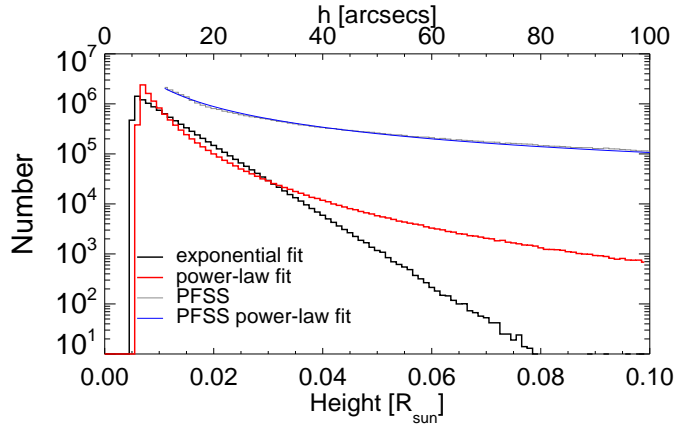}}
\caption{
The height distribution models compared.  The fitted exponential (black) and power-law (red) models ($\alpha = 3.1\pm0.1$) are shown.  The two models diverge significantly above 0.05$R_{\odot}$.  The  {\it PFSS} loop height distribution is 
also shown along with a power-law fit ($\alpha_{PFSS} = -1.4\pm0.12$).  The {\it PFSS} loop distribution has a much harder distribution than the measured distribution which may be due to the energization process of the corona. 
}\label{fig:height_distribution} 
\end{figure}

Using the standard {\it PFSS} software\footnote{see \url{http://www.lmsal.com/~derosa/pfsspack}} developed by M.\ DeRosa \citep[see][]{schrijver2003} and distributed with SolarSoft, the {\it PFSS} height model was calculated from 0.01 to 2.0$R_{\odot}$.  In order to get an accurate representative corona, the {\it PFSS} model was calculated every thirty days from March 2005 to March 2007, the time range of the RHESSI microflare list. Magnetic field lines were traced for each date and the maximum height for every (closed) loop was determined. From this the {\it PFSS} loop height distribution was then determined and each was fit with a power-law.  The distribution shown here is the average for every date.  The power-law index, $\alpha_{PFSS}$, of the average distribution was found to be $-1.4\pm0.12$.  The error in the power-law index represents the standard deviation of the set of fit values for individual dates.  The distribution of {\it PFSS} loop-top heights can be seen in Figure~\ref{fig:height_distribution} compared with the power-law and exponential fits to the microflare height distribution.  The {\it PFSS} height distribution was found to be much less steep than the power-law fit to the data with a difference in power-law index of $1.8\pm0.16$ which implies that the ensemble of potential loops contain more large loops than the observed distribution.  A soft loop height distribution may be a signature of the energization process of the corona.

\section{Conclusions}
\label{sec:conclusion}

Taking advantage of the large dataset provided by the RHESSI microflare list, the height 
distribution of HXR microflares was determined.  The distribution of HXR (4-10 keV) source heights was found to be well fit by an exponential distribution with a scale height of $6.1\pm0.3$~Mm.  A power-law height distribution with a negative power-law 
index of $3.1\pm0.1$ was also found to be consistent with the data though a worse fit.  The minimum observable height due to partially occulted sources was found to be $5.7\pm0.3$~Mm.  The value of the heights found here suggests that they are thermal loop sources.  Thermal loops are frequently assumed to be semicircular loops.  The center of mass of a semicircular loop is located 36\% below the loop-top therefore the minimum observable loop-top height may be as low as 7.7~Mm.  In practice, RHESSI images of loops do not usually show the base of the loop, therefore it is likely that this is an overestimate.

The height distributions were compared to the loop height distribution as determined by a 
potential field model ({\it PFSS}).  It was found that the measured flare height distribution was much steeper than the {\it 
PFSS} magnetic loop heights (difference in the power-law index of \appr1.8) meaning the ensemble of potential loops contain many more large loops compared to the height distribution found here.  Observations of rising post-flare loops in EUV suggest 
that the flare loop distribution may relax to the {\it PFSS} distribution as energy is dissipated.  In order for the observed
distribution to transition to the potential distribution small flare loops must rise faster than large flare loops.  It must also be
true that an energized corona suppresses large loops.

The height distribution for different energies (4--10~keV, 10--15~keV, 15--30~keV) were found to agree within statistical 
uncertainties.  More sensitive analysis showed that the difference in the average height were small, $
\Delta h_{4-10,10-15\ \mathrm{keV}} = 150\pm25$~km, $\Delta h_{10-15,15-30\ \mathrm{keV}} = 270\pm51$~km,
where lower energies are at greater heights.  This suggests that emission in each energy range should be interpreted as from thermal flare loops.

The minimum observable height found, $h_{0} = 6.3 \pm 0.3$~Mm, introduced by partially-occulted flares, suggests that investigations of flares at the limb should be mindful of this limitation whereby limb sources may appear at higher altitudes if they are partially behind the limb.  Such events should be detectable by the fact that images with different detectors will show positions which move away from the limb as a function of resolution.  It should also be possible to detect partially-occulted flares through the visibilities in the highest resolution detectors which should have unusually large amplitudes for directions parallel to the limb.  

%

%

%
\begin{acks}
This work was partially supported under NASA grant NAS5-98033 and NNM05ZA12H.  The authors would like to thank the referee for input which has greatly improved this paper.
\end{acks}

%
%
%

\bibliographystyle{spr-mp-sola-cnd} 
\bibliography{my_refs}  
\end{article} 
\end{document}